\documentclass[usegraphicx,useAMS,usenatbib,fleqn]{mn2e}
\include{latex_macros}
\usepackage{float}

\usepackage{times}
\usepackage{amsmath}
\usepackage{amsfonts}
\title {The effect of starspots on the ages of low-mass stars determined from the lithium depletion boundary}

\author[R. J. Jackson and R. D. Jeffries]
  {R. J.~Jackson and R. D.~Jeffries\\
  Astrophysics Group, Research Institute for the Environment, Physical
  Sciences and Applied Mathematics, Keele University, \\ Keele, 
      Staffordshire ST5 5BG
}
\date{In press}

\pagerange{\pageref{firstpage}--\pageref{lastpage}} \pubyear{2013}

\def\LaTeX{L\kern-.36em\raise.3ex\hbox{a}\kern-.15em
    T\kern-.1667em\lower.7ex\hbox{E}\kern-.125emX}

\begin{document}
\label{firstpage}
\maketitle

\begin{abstract}
In a coeval group of low-mass stars, the luminosity of the sharp
transition between stars that retain their initial lithium and those at
slightly higher masses in which Li has been depleted by nuclear
reactions, the lithium depletion
boundary (LDB), has been advanced as an almost model-independent means
of establishing an age scale for young stars. Here we construct
polytropic models of contracting pre-main sequence stars (PMS) that
have cool, magnetic starspots blocking a fraction $\beta$ of their
photospheric flux. Starspots slow the descent along Hayashi tracks,
leading to lower core temperatures and less Li destruction 
at a given mass and age. The age, $\tau_{\rm LDB}$,
determined from the luminosity of the LDB, $L_{\rm LDB}$, 
is increased by a factor
$(1-\beta)^{-E}$ compared to that inferred from unspotted models, where
$E \simeq 1 + d\log \tau_{\rm LDB}/d\log L_{\rm LDB}$ and has a value
$\sim 0.5$ at ages $<80$\,Myr, decreasing to $\sim 0.3$ for older
stars. Spotted stars have virtually the same relationship between
$K$-band bolometric correction and colour as unspotted stars, so this
relationship applies equally to ages inferred from the absolute $K$
magnitude of the LDB. Low-mass PMS stars do have starspots, but
the appropriate value of $\beta$ is highly uncertain with a probable
range of $0.1 < \beta < 0.4$. For the smaller $\beta$ values our result
suggests a modest systematic increase in LDB ages that is comparable
with the maximum levels of theoretical uncertainty previously claimed
for the technique. The largest $\beta$ values would however increase LDB ages
by 20--30 per cent and demand a re-evaluation of other age estimation
techniques calibrated using LDB ages.

\end{abstract}


\section{Introduction}

During their approach to the main sequence, the cores of contracting
low-mass ($\leq 0.5\,M_{\odot}$) pre-main sequence (PMS) stars reach
temperatures sufficient to ignite lithium. Rapid mixing in these
fully-convective stars then leads to complete Li destruction throughout
the star on a time scale that is short compared with their contraction time
scales (e.g. Basri, Marcy \& Graham 1996; Rebolo et al. 1996; Bildsten
et al. 1997). The time it takes to reach this Li-burning threshold
increases with decreasing stellar mass.  If a set of coeval low-mass
stars in a young cluster are spectroscopically examined for the
presence of photospheric Li, there is an abrupt transition from stars
that have depleted all their initial Li to
stars of only slightly lower luminosity (and mass) that retain their
initial Li.  The luminosity, $L_{\rm LDB}$,
at this ``Lithium Depletion Boundary'' (LDB) can therefore be
used to estimate the age of the cluster, $\tau_{\rm LDB}$ 
(Stauffer, Schultz \& Kirkpatrick 1998; Stauffer et al. 1999;
Jeffries \& Oliveira 2005).
\nocite{Basri1996a}
\nocite{Rebolo1996a}
\nocite{Bildsten1997a}
\nocite{Stauffer1998a}
\nocite{Stauffer1999a}
\nocite{Jeffries2005a}

$\tau_{\rm LDB}$ is approximately proportional to $L_{\rm LDB}^{-1/2}$
(see section 2.3), so with typical uncertainties in distances and
bolometric corrections, LDB age estimates can have precisions of 5--10
per cent. Of greater importance is that relatively few assumptions need
be made in order to calculate $L_{\rm LDB}$ as a function
of age; analytical calculations and evolutionary models incorporating
differing treatments of opacities, equations of state and convection
produce age estimates that differ by less than 15 per cent (Ushomirsky
et al. 1998; Jeffries \& Naylor 2001). Burke, Pinsonneault \& Sills
(2004) conducted a series of calculations, altering the physical inputs
of their evolutionary models within plausible bounds, finding
"theoretical uncertainties" in LDB ages that were just 3--8 per cent
depending on the stellar mass at the LDB.

\nocite{Ushomirsky1998a}
\nocite{Jeffries2001b}
\nocite{Burke2004a}

The theoretical robustness of LDB age estimates compares favourably
with competing methods of cluster age determination (e.g. Soderblom
2010; Soderblom et al. 2013). Ages determined from isochrone-fitting to
high-mass stars, as they leave the ZAMS and turn off the main-sequence,
depend sensitively on the effects of rotation and the assumed levels of
convective overshoot in the core (e.g. Maeder \& Meynet 1989; Meynet \&
Maeder 2000; Ekstr\"om et al. 2012). Isochrone fitting to low-mass
PMS stars is subject to significant uncertainties in the efficiency of
convection and the adopted outer atmospheres (e.g. D'Antona \&
Mazzitelli 1994; Baraffe et al. 1998), and there are further
uncertainties introduced in transforming observables such as spectral
type into effective temperatures.  Resultant age uncertainties can be
factors of two, even given a large set of coeval stars at a range of
masses (e.g. Hillenbrand, Bauermeister \& White 2008; Hillenbrand
2009). For these reasons, a review of techniques for estimating young
stellar ages by Soderblom et al. (2013) concluded that the LDB method
provides the most reliable means of setting the absolute age scale of
PMS stars.

\nocite{Soderblom2010a}
\nocite{Soderblom2013a}
\nocite{Maeder1989a}
\nocite{Meynet2000a}
\nocite{Ekstrom2012a}
\nocite{Dantona1994a}
\nocite{Baraffe1998a}
\nocite{Hillenbrand2008a}
\nocite{Hillenbrand2009a}

LDB age determinations now exist for 9 clusters and associations (see
Soderblom et al. 2013 and references therein), ranging in age from
20--130\,Myr. At ages $< 20$\,Myr, the extent of the superadiabatic
envelope leads to rapidly growing model dependencies on the adopted opacities
and convection treatment (Burke et al. 2004). At ages $\geq 150$ \,Myr, the
sensitivity of the technique declines, but a greater limitation is that
$L_{\rm LDB}$ becomes so small that spectroscopy in even nearby
clusters becomes impractical.

Despite the optimism that LDB ages are robust to theoretical
uncertainties, the assumptions made and the effects neglected by {\it
  all} evolutionary models could still lead to systematic errors. An
important example is the neglect of the effects of rotation and
magnetic activity -- both of which are manifestly present in low-mass
PMS stars. Whilst the effects of rotation on the hydrostatic structure
of low-mass PMS stars and their Li depletion are likely to be small
(e.g. Mendes, D'Antona \& Mazzitelli 1999; Burke et al. 2004), the
effects of the consequent dynamo-generated magnetic fields may be more
significant (e.g. Ventura et al. 1998; D'Antona, Ventura \& Mazzitelli
2000). There is growing observational evidence from fast-rotating,
magnetically active stars in tidally-locked eclipsing binaries, young
clusters and the field, that magnetic activity may increase the radii of
low-mass stars (Lopez-Morales 2007; Morales, Ribas \& Jordi 2008;
Jackson, Jeffries \& Maxted 2009; Stassun et al. 2012). The mechanism
by which it does so is still unclear; but could include the magnetic
stabilisation of the star against convection (Gough \& Tayler 1966;
Moss 1968; Mullan \& Macdonald 2001; Feiden \& Chaboyer 2013), a
reduction in convective efficiency due to a turbulent dynamo (Feiden \&
Chaboyer 2014) or the blocking of emergent flux by dark, magnetic
starspots (Spruit 1982; Spruit \& Weiss 1986; Jackson \& Jeffries
2014). If PMS stellar radii are increased by magnetic activity, they could have
lower central temperatures and hence less Li depletion at a given
age. Using an arbitrary lowering of the mixing length parameter to
simulate a reduction in convective efficiency, Somers \& Pinsonneault
(2014) show that spreads in Li abundance seen in the G/K stars of
young clusters can be explained by varying levels of magnetic activity
connected with a spread in rotation rates. They also claim that a
similar increase in radius associated with magnetic fields in lower
mass stars 
would delay the onset of Li burning and that current 
LDB ages may be underestimates.

\nocite{Spruit1982a}
\nocite{Moss1968a}
\nocite{Mendes1999a}
\nocite{LopezMorales2007a}
\nocite{Morales2008a}
\nocite{Jackson2009a}
\nocite{Stassun2012a}
\nocite{Mullan2001a}
\nocite{Gough1966a}
\nocite{Feiden2013a}
\nocite{Feiden2014a}
\nocite{Jackson2014a}
\nocite{Spruit1986a}
\nocite{Somers2014a}
\nocite{Ventura1998a}
\nocite{Dantona2000a}

In Jackson \& Jeffries (2014) we investigated how the presence of dark,
magnetically induced starspots influences the evolution and radii of
low-mass PMS stars. Using a polytropic model, we showed that the radii
of spotted,
fully convective PMS stars are increased by a factor $\simeq (1 -
\beta)^{0.45}$ compared with unspotted stars of the same luminosity,
where $\beta$ is the fraction of photospheric flux blocked by spots. In
this contribution we use a similar approach to explore how
starspots will affect ages determined from the LDB. We find that
for plausible levels of spot coverage on active, low-mass PMS stars,
that current LDB ages are {\it systematically} underestimated by a
factor comparable to, or larger than, 
the maximum levels of theoretical uncertainty
previously claimed for the technique. 

The plan of the paper is as follows: section 2 describes how we adapt
and calibrate a simple polytropic model to predict the time at which Li
is depleted by a given fraction in fully convective low-mass
stars. Provided that the critical temperature at which Li burning commences
and the properties of the atmospheric opacities do not change rapidly
with mass, this leads to a relatively simple expression for the
increase in age implied for spotted stars at a given $L_{\rm LDB}$. In
section 3 we then consider the properties of the starspots and how
these alter bolometric corrections and the relationship between
absolute magnitude of the LDB and age. In section 4 we discuss the
implications of the results and in particular, the consequences of
older LDB ages for evolutionary models and the calibration of
isochronal ages in the HR diagram.

\section{Effect of starspots on the age of lithium depletion}

The presence of dark starspots on the photosphere will slow the
contraction of PMS stars, leading to larger radii at a given luminosity
(Jackson \& Jeffries 2014). This will delay the onset of lithium
depletion and hence in a cluster of coeval stars, alter the
relationship between $L_{\rm LDB}$ and $\tau_{\rm LDB}$. 
We evaluate this as follows:

A set of model evolutionary tracks are analysed to determine
both the age, $t_{\rm Li}$ and radius, $R_{\rm Li}$ as a function of
mass, $M$,  and luminosity, $L$, at which low mass stars achieve a
given level of lithium depletion (typically, 95 or 99 percent thresholds
are used by observers, e.g. Jeffries \& Oliveira 2005).

A polytropic model is introduced to represent fully convective low-mass 
stars as they descend Hayashi tracks towards the LDB. Constants
defining the surface opacity of the polytropic model are chosen to fit
the results of published (unspotted) 
evolutionary tracks as a function of $M$ in the
Hertzsprung-Russell (HR) diagram.

The polytropic model is used to determine the effect of starspots
on $R$ and $L$ as a function of $M$, age and starspot coverage. Under
the assumption that similar constants characterise the polytropic models
for spotted and unspotted stars at the LDB, and that Li depletion occurs
at the same value of $R/M$ in spotted and unspotted stars, we derive a
simple analytic expression for the relative change in the LDB age of
spotted stars as a function of spot coverage.

\begin{figure*}
	\centering
	\begin{minipage}[t]{0.95\textwidth}
	\centering
	\includegraphics[width = 150mm]{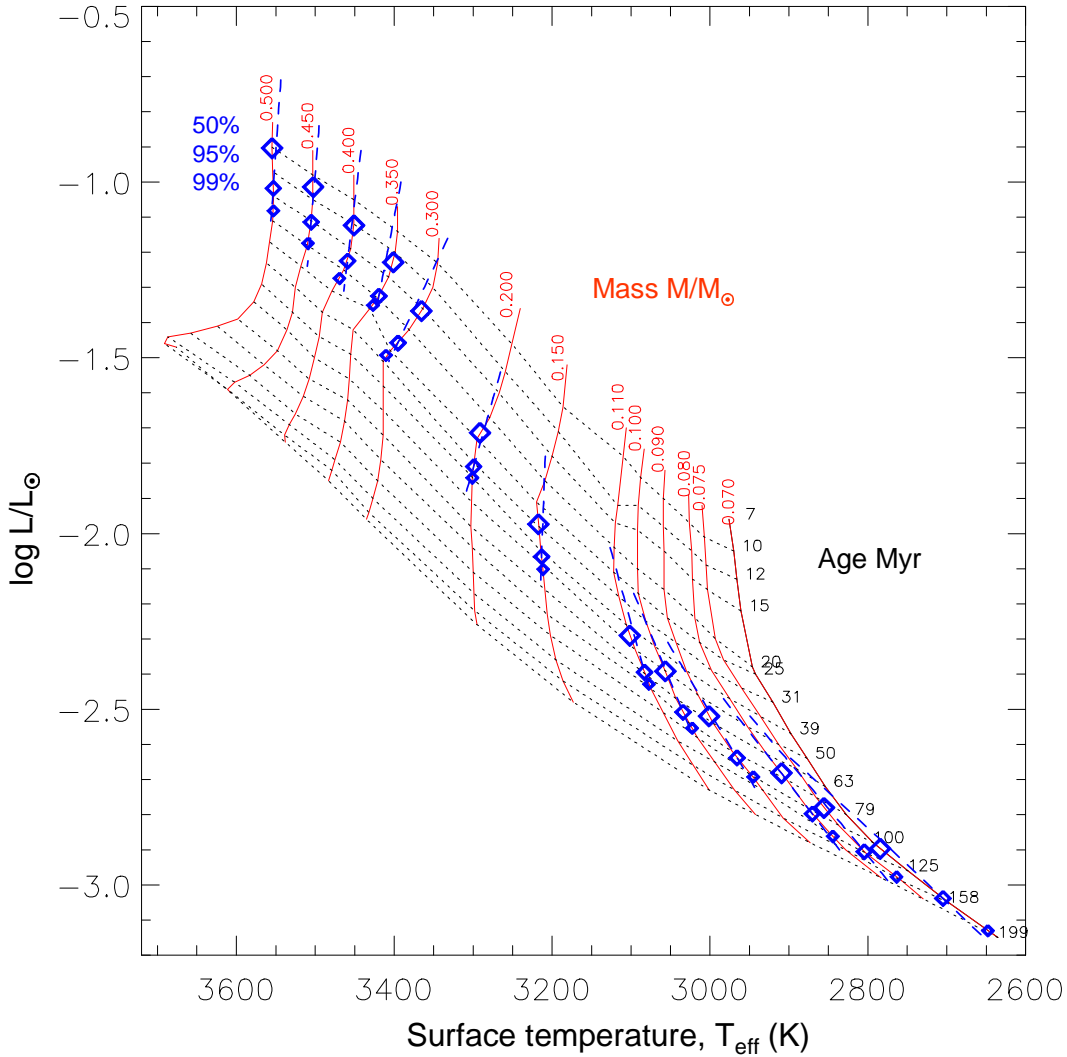}
	\end{minipage}
	\caption{Model isochrones and evolutionary tracks for young low
          mass stars from the Baraffe et al. (1998) models of stellar evolution
          ([M/H]=0.0, Y=0.275, $L_{\rm mix}=H_p$). Diamonds of
          decreasing size along each evolutionary track indicate
          the surface temperature and luminosity at the age of 50, 95
          and 99 percent lithium depletion. Blue dashed lines show
          evolutionary tracks predicted by a polytropic model of a
          fully convective
          unspotted star in the Hayashi zone with parameters fitted to
          match the published numerical models as they approach the 95 percent Li
          depletion age (see section 2.2).}
\label{fig1}
\end{figure*}

\begin{figure*}
	\centering
	\begin{minipage}[t]{0.95\textwidth}
	\centering
	\includegraphics[width = 150mm]{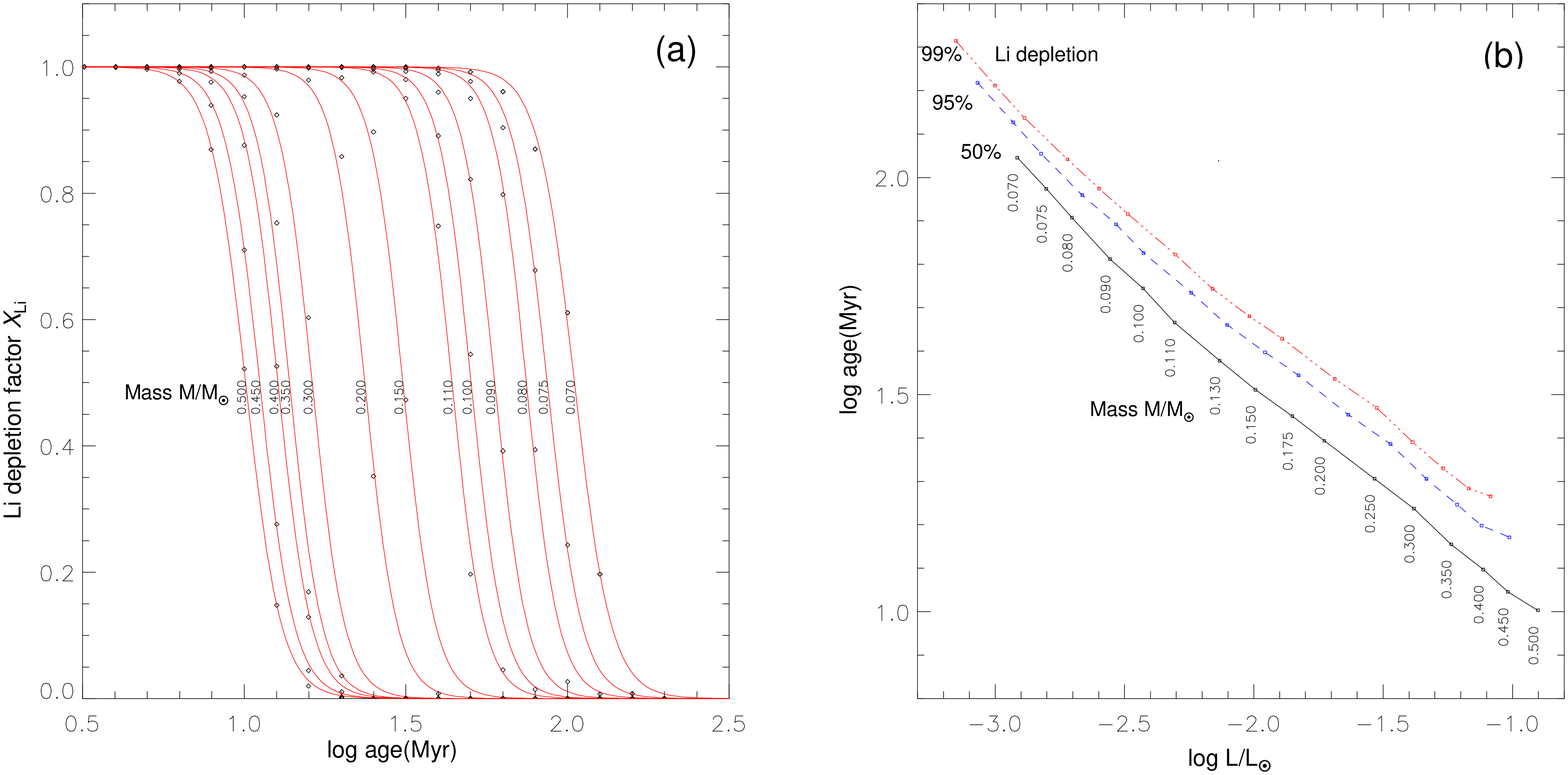}
	\end{minipage}
	\caption{Variation of the lithium depletion factor, $X_{Li}$
          with age for low mass stars. Crosses in the left hand plot
          show tabulated values of $X_{\rm Li}$ for the BCAH98 model of
          stellar evolution ([M/H]=0.0, Y=0.275,
          $L_{\rm mix}=H_p$). Curves show the function $X_{\rm Li} =
          \tfrac{1}{2}[1-\tanh ((t-t_{50})/\tau_{\rm Li} )]$ fitted at
          each mass (where $t_{50}$ is the 50 percent Li depletion and
          $\tau_{\rm Li}$ the time constant for decay). The right hand
          plots show the 50, 95 and 99 per cent Li depletion age as a
          function of luminosity using these fitted parameters.}       
\label{fig2}
\end{figure*}

\subsection{Masses and radii of stars at the LDB}
Fig.~1 shows evolutionary tracks and associated isochrones for low
mass stars ($0.07<M/M_{\odot}<0.5$). These are outputs plotted from the
BCAH98  theoretical isochrones (Baraffe et al. 1998)
for metallicity [M/H]=0.0, helium mass fraction Y=0.275 and a convective
mixing length of one pressure scale height ($L_{\rm mix}=H_p$). The tabulated model data include a lithium depletion factor, $X_{\rm
  Li}$, as a function of mass and age, representing the fraction of the
initial photospheric Li remaining. These data are interpolated by
fitting $X_{\rm Li} = \tfrac{1}{2}[1-\tanh ((t-t_{50})/\tau_{\rm Li}
  )]$ to the tabulated isochrones, where the constant $t_{50}$
represents the (mass-dependent) epoch of 50 per cent Li depletion and $\tau_{\rm Li}$
is a time constant for Li destruction (see Fig.~2a). 
These results are used to calculate the age at which 50, 95
and 99 per cent lithium depletion occurs as a function of mass. These
points are marked on the individual evolutionary tracks in Fig. 1 and
are plotted as a function of luminosity in Fig.2b. The time scale over
which Li is burned is indicated by the gap between the curves in
Fig.~2a. It is quite uniform, although becomes slightly shorter (in
terms of a logarithmic age increment) at lower masses and luminosities.

\nocite{Allard2011a}
\nocite{Baraffe1998a}
\nocite{Asplund2009a}

Extrapolating these results to stars with varying levels of spot
coverage requires an additional relation representing the dependence of
the rate of lithium depletion on the temperature, $T_c$, and density,
$\rho_c$, at the stellar core. The reaction rate for lithium depletion
is extremely temperature sensitive ($\propto T_{c}^{20}$) with a much
weaker dependence on density. The resultant narrow range of burning
temperatures allows $T_c$ and $R$ at the time of Li
depletion to be approximated by power laws, at least for a
non-degenerate star (e.g. Bildsten et al. 1997). For fully convective,
non degenerate, PMS stars the core properties scale approximately as
$T_c \propto M/R$ and $\rho_c \propto M/R^3$. Hence the ratio of radius
to mass for a given level of 
lithium depletion is expected to be
nearly invariant, or at least only a slowly varying function of mass.

Fig.~3 shows the critical radius to mass ratio, $[R/M]_{\rm Li}$ at
which 50, 95 and 99 per cent of Li is depleted as a function of surface
temperature and mass, determined from the BCAH98 models.
For $M > 0.11\,M_{\odot}$, $[R/M]_{\rm Li}$ only changes 
as $\sim
M^{-0.1 }$, which is consistent with the analytic prediction of Bildsten
et al. (1997) that the radius at the LDB for non-degenerate stars
varies as $M^{7/8}\mu^{3/4}T_{\rm eff}^{-1/6}$, where $\mu$ is the mean
atomic mass and the effective temperature of Hayashi tracks 
varies little with mass.
For stars (and brown dwarfs) with $M \leq 0.11\,M_{\odot}$, $[R/M]_{\rm
  Li}$ peaks and decreases slowly towards lower masses.
This is presumably due to the onset of electron degeneracy and the consequent
increase in the effective atomic mass, $\mu_{\rm eff}$, that is defined in
terms of the ratio of gas temperature to pressure (see section 2.2 and
Ushomirsky et al. 1998). 

The fact that
$[R/M]_{\rm Li}$ does not vary strongly with $M$ or
$T_{\rm eff}$ over the ranges relevant to low-mass PMS stars, in both the non-degenerate
and partially degenerate regimes, is what allows us to obtain
analytic expressions for the effects of starspots on LDB ages.


\subsection{Polytropic models as function of mass}
Low-mass stars over the mass range shown in Fig. 1 are predicted to
remain fully convective at least until the end of any Li burning
(D'Antona \& Mazzitelli 1994). Fully and efficiently convective stars
can be represented by a polytropic model with the equation of state of
an ideal gas, $P= N_a k_b \rho T/\mu$, where $N_a$ is Avagadro's number
and $k_b$ is Boltzmann's constant. For this case the adiabatic pressure
relation is $P\propto \rho^{5/3}$ and the polytropic index $n=3/2$. This 
relation is valid even when the core of the star becomes partially
degenerate at the lower mass and older extents of the parameter ranges
considered in this paper. In an isentropic star the degree
of electron-degeneracy is constant (Stevenson 1991). Hence the effect
of degeneracy is simply to increase the effective mean molecular
mass, $\mu_{\rm eff}$, thus reducing the temperature through the star by a
constant factor $\mu/\mu_{\rm eff}$. The adiabatic exponent remains
unchanged and the star is described by a $n=3/2$ polytrope,
independent of the degree of degeneracy (Hayashi \& Nakano 1963;
Ushomirsky et al. 1998).

\nocite{Hayashi1963a}
\nocite{Stevenson1991a}

\begin{figure}
	\centering
	\includegraphics[width = 75mm]{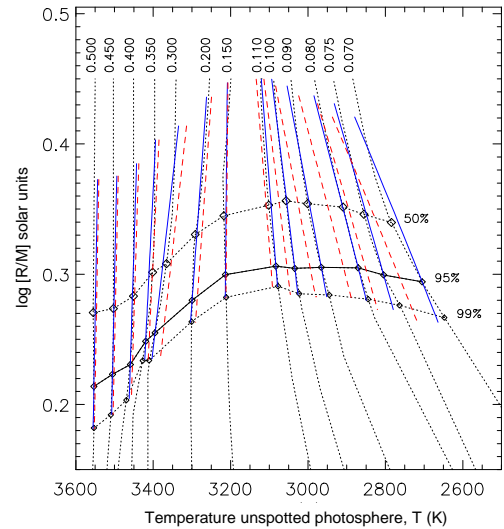}
	\caption{The critical radius/mass ratio,
          $[R/M]_{Li}$ at which 50, 95 and 99 per cent Li depletion
          occurs (marked with diamonds of decreasing size) 
          shown on evolutionary tracks in the $R/M$
          versus temperature plane for a range of labeled stellar masses. 
          Dotted lines show evolutionary tracks for (unspotted) BCAH98
          models, where the x-axis shows the temperature of the unspotted photosphere ($T_{eff}$
          for an unspotted star). Blue lines show tracks predicted using a polytropic model of an
          unspotted star in the Hayashi zone with parameters fitted to
          match the evolutionary tracks in the HR diagram (see
          Fig.~1). Red dashed lines show the displacement of the tracks
          resulting from 30 percent spot coverage of dark starspots
          ($\beta=0.3$, see section 2.4).}
\label{fig3}
\end{figure}
\begin{figure}
	\centering
	\includegraphics[width = 78mm]{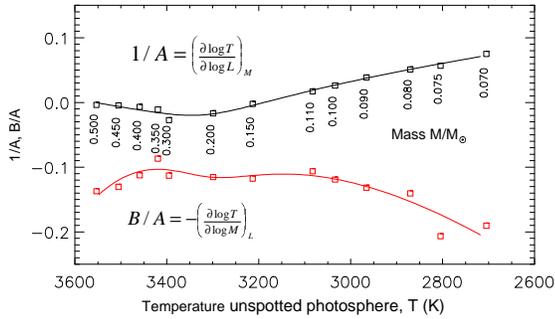}
	\caption{The variation of the exponents A and B (defined in
          equation~1, plotted as
          $1/A$ and $B/A$ respectively) defining the
          polytropic model of a star in the Hayashi zone as a function of mass and
          photospheric temperature (at the epoch of 95 per cent Li
          depletion). Squares give values at individual masses. The curves
          are fifth order polynomial fits.}
\label{fig4}
\end{figure}

\begin{figure}
	\centering
	\includegraphics[width = 78mm]{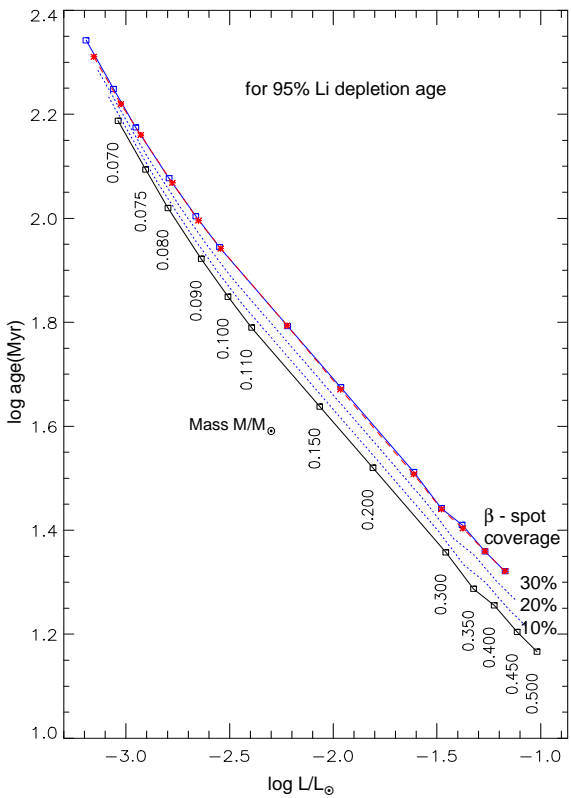}
	\caption{The effect of starspots on the Li depletion age as a
          function of luminosity. The lower solid line shows the 95
          percent Li depletion age for an unspotted star. Blue dotted
          and solid lines show the effect of increasing levels of spot
          coverage ($\beta$=0.1,\,0.2 and 0.3) where, at a given mass
          (marked with fiducial squares in the lower and upper curves),
          the Li depletion age is increased by a factor
          $(1-\beta)^{-1}$ and the luminosity is reduced by a factor
          $(1-\beta)$. The red line (and mass points marked as red crosses) show results for $\beta$=0.3 taking account of the small changes in parameters defining the polytropic model caused by starspot coverage (see section 2.4).}
	
\label{fig5}
\end{figure}

The first step is to fit a polytropic model to the 
evolutionary tracks of the {\it unspotted} PMS stars in Fig.1
as they approach the epoch of Li depletion. A fully convective, contracting 
polytropic PMS star follows a (Hayashi) track in the HR diagram given by
\begin{equation}
\frac{L}{L_{\odot}}=C^{-A}\left(\frac{M}{M_{\odot}}\right)^B\left(\frac{T}{T_{\odot}}\right)^A,
\label{eqn1} 
\end{equation}
where $T$ is the photospheric temperature and $A$,$B$ and $C$ are
constants calculated from the polytropic index,
$n$ and the indices $a$ and $b$ used to describe how the Rosseland mean
opacity depends on the temperature and density at the photosphere
($\kappa=\kappa_0\rho^a T^b$, see Prialnik 2000; Jackson \& Jeffries 2014).
In practice $\kappa$ varies in a complex manner with
temperature such that it cannot be described by single values of $a$
and $b$ over the full range of temperatures and masses shown in
Fig. 1. Instead, local values of $A$ and $B$ are
determined as a function of mass from the  
evolutionary tracks. With $n$ fixed at $3/2$, 
the value of $A$ is determined from the slope of
the evolutionary tracks at constant mass ($1/A =\tfrac{\partial \log
  T}{\partial \log L}$). This slope changes with time, so we
restrict the fit to the 0.7~dex of time immediately preceding the epoch
of 95 per cent lithium depletion, $t_{95}$ (see Fig. 1). The rationale
for this is that it will only be the presence of spots and the
consequent structure of the star during this period that affects
Li depletion.
The value of $B$ is
determined in a similar way from the spacing between the evolutionary tracks ($B/A
=-\tfrac{\partial \log T}{\partial \log M}$) as a function of $M$ and
$T_{\rm eff}$ (at
$t_{95}$). The constant $C$ is determined from the luminosity at
$t_{95}$. The resulting values of $1/A$ and $B/A$ are shown as a function
of temperature and mass in Fig.~4. Tracks based on these simple polytropic models
(dashed lines in Fig.~1) are able to reproduce the results of
the numerical calculations for stars approaching the Li depletion epoch. 

\subsection{The effect of starspots on $R$ and $L$}
Jackson \& Jeffries (2014) used a similar polytropic model to investigate the effect of dark
photospheric spots on the evolution of low mass stars as they descend
Hayashi tracks, where the luminosity of the star is taken to result
solely from the release of gravitational energy. When spots form on the
stellar surface their immediate effect is to reduce $L$ by
a factor ($1-\beta$), with no change in radius or temperature of
the unspotted photosphere, $T_u$. Here $\beta$ is the equivalent filling
factor of completely dark starspots that would produce the same
reduction in $L$ as the actual coverage of dark starspots at the
actual (non zero) spot temperature. The subsequent effect of starspots is
to reduce the rate of descent along the Hayashi track by a factor
$(1-\beta )$. In the long-term there is an increase in $R$
at a given age $t$, relative to the radius of unspotted stars of
similar mass. Jackson \& Jeffries showed that provided starpots were formed more than
about 1~dex previously in $\log t$, then\footnote{Given that evidence
  for spots is plentiful on stars as young as $\sim 1$\,Myr
  (e.g. Herbst et al. 2002), this assumption seems reasonable at ages
  $>10$\,Myr.}
\begin{equation}
\frac{R}{R_{\odot}}=[Zt( 1-\beta)] ^{\frac{A-4}{4-3A}} 
\left( \frac{M}{M_{\odot}} \right)^{\frac{8-2A-4B}{4-3A}}\, ,
\label{eqn2} 
\end{equation}
\begin{equation}\frac{L}{L_{\odot}}=(1-\beta)C^{\frac{-4A}{4-A}}[Zt(1-\beta)] ^{\frac{2A}{4-3A}} 
\left( \frac{M}{M_{\odot}} \right) ^{\frac{4B-4A}{4-3A}}\, ,
\label{eqn3} 
\end{equation}
\begin{displaymath}
{\rm where}\ \ Z = C^{\frac{-4A}{4-A}}\tfrac{(10-2n)(4-3A)}{3(4-A)}\left(\frac{L_{\odot}R_{\odot}}{GM_{\odot}^2}\right)\,.
\end{displaymath}

\begin{figure}
	\centering
	\includegraphics[width = 75mm]{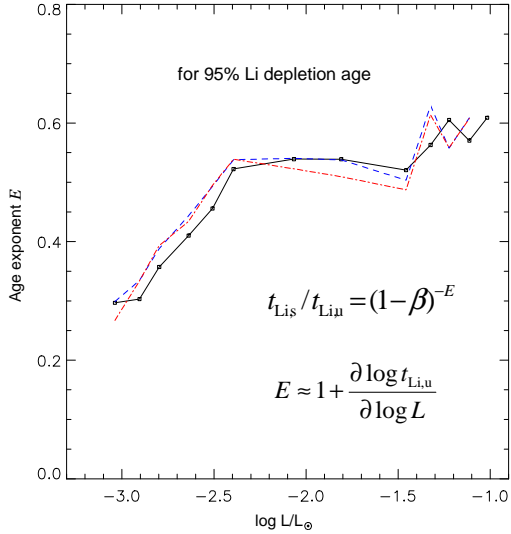}
	\caption{Variation of the exponent, $E$ defining the increase
          in Li depletion age of a spotted star, $t_{\rm Li,s}$
          relative to that of an unspotted star, $t_{\rm Li,u}$ of similar
          luminosity. The solid black line shows $E$ calculated from
          the slope of the 95 percent Li depletion age curve
          (Fig.~2). The blue dashed line shows the exponent calculated
          from the curves in Fig.~5 where, at a given mass, the Li
          depletion age is increased by a factor $(1-\beta)^{-1}$ and
          the luminosity is reduced by a factor $(1-\beta)$. The red
          dot-dashed line shows results for $\beta$=0.3 taking account
          of the small changes in parameters defining the polytropic
          model caused by starspot coverage (see section 2.4).}
\label{fig6}
\end{figure}

\nocite{Herbst2002a}

These expressions can be used as follows to determine the change in the Li
depletion age of a spotted star, $t_{\rm Li,s}$, relative to that of an
unspotted star, $t_{\rm Li,u}$.
\begin{enumerate}
\item Consider an unspotted star of mass $M$ requiring a critical value of
  $[R/M]_{\rm Li}$ to achieve a specified level of Li depletion,
  which is reached at an age $t_{\rm Li,u}$ (see Fig.3). From equation~\ref{eqn2} the same value of
  $[R/M]_{\rm Li}$ is reached for a spotted star at an age of
  $(1-\beta)^{-1} t_{\rm Li,u}$. Hence {\it at a fixed mass}, $[t_{\rm
      Li,s}/t_{\rm Li,u}]_M = (1 - \beta)^{-1}$, corresponding to a 
  vertical shift of a mass point upwards in Fig~2b. 
  This assumes
  that the photospheric temperature and density of the spotted star are
  not changed enough to alter the values of $A$ and
  $B$ sufficiently to affect the comparison, and that $[R/M]_{\rm Li}$
  is the same for spotted and unspotted stars.

\item However, the corresponding change in luminosity at fixed mass between an
   unspotted star, $L_{\rm u}$ at age, $t_{\rm Li,u}$ and a spotted star,
   $L_{\rm s}$ at age $t_{\rm Li,u} (1-\beta )^{-1}$ is given by
   equation~\ref{eqn3} as $[L_{\rm s}/L_{\rm u}]_M = (1-\beta )$. This corresponds 
   to a horizontal shift to the left in Fig.~2b, 
   assuming again that the indices in equation~\ref{eqn3} that depend on $A$
   and $B$ do not change significantly for the spotted star.


\end{enumerate}

Fig.~5 shows a logarithmic plot of $t_{\rm Li}$ versus
luminosity using a 95 per cent Li depletion criterion. The lower
solid line shows the relation for
an unspotted star interpolated from the BCAH98 models shown in
Fig.~2. The blue dashed and solid lines show the effect of 10, 20 and
30 per cent spot coverage where, for each mass point, the age has been 
increased by a factor $(1-\beta)^{-1}$ and the luminosity reduced by a factor
$(1-\beta)$. 

These two effects can be combined into one simple approximation that
$t_{\rm Li,s} \simeq t_{\rm Li,u} (1 - \beta)^{-E}$, where $E-1$ 
  is the gradient of the original (unspotted) $t_{\rm Li}$ versus
  $L$ relationship. i.e. $E \simeq 1+\partial \log t_{\rm
      Li,u}/\partial \log L$. This approximation is true for small $\beta$
  and remains accurate for larger $\beta$ provided $\log
  t_{\rm Li,u}$ varies linearly with $\log L$. Fig.~6 compares the
  value of $E$ determined from the slope of the unspotted Li depletion age
  curve in Fig.~5 with the effective values of $E$ found
  from the positions of the age curves for spotted stars shown in Fig.~5 
  that were shifted according to the
  two-step process described above. Results differ by only 0.03 rms over
  the mass range 0.07$<M/M_{\odot}<$0.4. $E \simeq 0.5$ over most of
  this mass range, so to first order we can say
  that {\it the LDB ages inferred from models of spotted stars are
  older than those inferred from standard models by a factor of} $\sim
  (1- \beta)^{-1/2}$, but by a slightly smaller 
  factor for older ($>80$\,Myr) clusters with
  $L_{\rm LDB} < 10^{-2.5} L_{\odot}$, where $E \sim 0.3$ (see Figs. 5 and 6).
  
 A further effect of starspots is to change the inferred
 mass of stars at the LDB. From the mass points in Fig.~5 it can be seen
 that starspots both increase the Li depletion age at a given luminosity and  
 increase the mass of the star that reaches its Li depletion age at
 this luminosity. From equations 2 and 3 the
 stellar mass at the LDB scales with spot coverage as $(M_s/M_u)_{\rm LDB} = (1
 - \beta)^{-(A-4)/(2A-4B)}$. As $A$ is large compared with $B$
 (see Fig.~4), $(M_s/M_u)_{\rm LDB} \simeq (1- \beta)^{-1/2}$.

\subsection{The effect of changes in polytropic constants}
In calculating the change in Li depletion age with $\beta$ it is
implicitly assumed that the constants defining the polytropic model of
a spotted star are the same as those of an unspotted star. The validity
of this approximation can be tested by evaluating 
$t_{\rm Li, s}$ taking account of the changes in
critical parameters with spot coverage. In order to do this it is
convenient to define the change in Li depletion age and luminosity in
terms of the temperature $T$ (defined as the temperature of the
spot-free region of the photosphere), the critical radius for Li
depletion, $[R/M]_{\rm Li}$ and the exponent $A$. From
equations 2 and 3 the change in Li depletion age at fixed mass due to
starspot coverage is actually
\begin{equation}
	\left(\frac{t_{\rm Li, s}}{t_{\rm Li, u}}\right)_M =
        \frac{1}{(1-\beta)} \frac{\tfrac{4-A_s}{4-3A_s}[R/M]_{\rm
            Li,s}^{-3} T_s^{-4}} {\tfrac{4-A_u}{4-3A_u}[R/M]_{\rm
            Li,u}^{-3} T_u^{-4}}\, ,
\end{equation}
where suffix $s$ denotes values of $A$, $[R/M]_{\rm Li}$ and $T$ for a
spotted star and suffix $u$ denotes values for an unspotted star. Using
the relation $L = (1-\beta)R^2 T^4$, the change in luminosity at the Li
depletion age is
\begin{equation}
	\left(\frac{L_{\rm Li, s}}{L_{\rm Li, u}}\right)_M = (1-\beta)
        \frac{[R/M]_{\rm Li,s}^{2} T_s^{4}} {[R/M]_{\rm Li,u}^{2}
          T_u^{4}}
\end{equation}

It was shown in Jackson \& Jeffries (2014) that the effect of starspots
is to offset the Hayashi track of the spotted star relative to that of
an unspotted star of the same mass in the HR diagram. The offset at
fixed mass and age depends on the slope of the Hayashi track, as
$(T_s/T_u)_{M,t} = (1-\beta)^{-2/(3A-4)}$ (see equation 9 of Jackson and
Jeffries). The change in $T$ produces small changes in $[R/M]_{\rm Li}$
and $A$. Fig. 4 shows the displacement of the polytropic tracks of
$[R/M]_{\rm Li}$ versus $T$ due to spot coverage which gives an upper bound for the
change in $[R/M]_{\rm Li}$ for each mass point. Similarly the change in
$A$ for a spotted star can be estimated by interpolating the polynomial
approximation of $1/A$ versus $\log T$ in Fig.~4.

Fig.~5 also shows an LDB-age relation for a heavily spotted star
($\beta$=0.3) taking account of this change in $T$ due to spot coverage
and the resulting changes in $[R/M]_{\rm Li}$ and $A$. The curve is
almost identical to the curve calculated assuming that $[R/M]_{\rm Li}$
and $A$ are fixed, although at low masses the equivalent mass points
are displaced parallel to the curve so the {\it mass} of a star at
$L_{\rm LDB}$ is slightly smaller than would otherwise be estimated. In
effect, the change of surface temperature at low masses reduces the age
of the spotted star (at fixed mass) but increases its luminosity, such
that the change of age at fixed luminosity is only slightly
modified. Fig.~6 compares the age exponent $E$ taking account of
changes in $T$, $([R/M]_{\rm Li}$ and $A$ with the results assuming
these parameters are fixed. For a high level of spot coverage
($\beta$=0.3) the age exponent differs by $<0.02$ rms over the mass
range 0.07$<M/M_{\odot}<$0.4 and therefore we judge this a negligible
source of uncertainty when estimating LDB ages.

\section{The effects of star spots on the bolometric correction}
\begin{figure}
	\centering \includegraphics[width = 75mm]{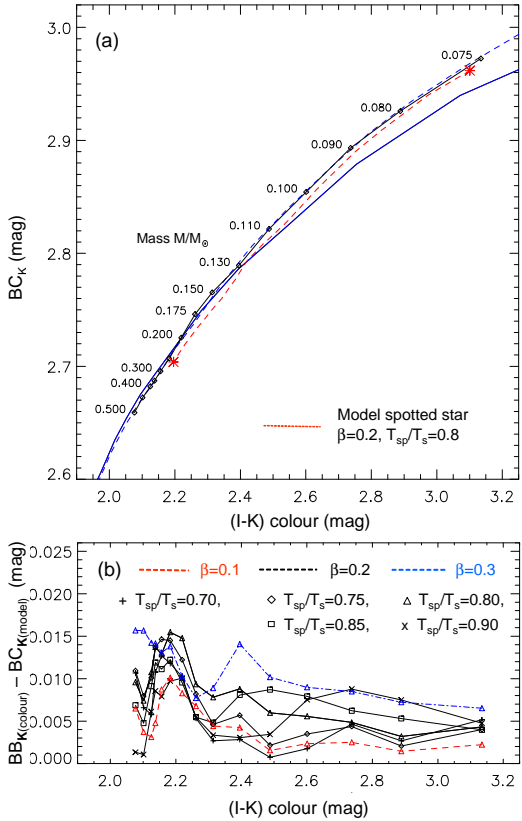}
	\caption{The effect of starspots on the BC$_K$ bolometric
          correction inferred from an $I-K$ colour. The upper plot
          shows BC$_K$ versus colour using BT-Settl model atmospheres
          (Allard et al. 2011). The blue solid and dashed lines show
          curves for unspotted stars at fixed ages of 1\,Gyr and 150\,Myr
          respectively. The solid black line shows BC$_K$ values at the epoch
          of 95 per cent Li depletion. The red dashed line shows the
          predicted values of BC$_K$ for spotted stars with $\beta$=0.2
          and a spot temperature contrast of 0.8. The lower plot show
          the difference in the value BC$_K$ if this evaluated from its
          colour (assuming there are no star spots present) compared to
          its value for a spotted star predicted using a
          two-temperature model. Results are shown for a range of spot
          temperature contrast ratios, and spot coverages $\beta$.}
\label{fig7}
\end{figure}
As luminosities and temperatures cannot be measured directly, the
observational route to inferring an LDB age involves estimating the
bolometric magnitude of the LDB using its apparent magnitude, along
with distance, extinction and either empirical or theoretical
relationships between bolometric correction and the colour or spectral
type of stars at the LDB (e.g. Jeffries \& Oliveira 2005; Binks \&
Jeffries 2014).  In this section we consider how the presence of
starspots affects the relationship between bolometric correction and
colour for spotted stars. We focus on $K$ magnitudes and bolometric
corrections versus $I-K$, as these are most frequently used data in the
LDB literature.
\nocite{Binks2014a}

\subsection{Colour transformations for unspotted stars}
Fig.~7a shows curves of the $K$-band bolometric correction, BC$_K$, as
a function of $I-K$ colour interpolated from the "BT-Settl/AGSS2009"
model which use synthetic colours and magnitudes derived from the
BT-Settl model atmospheres (Allard, Homeier \& Freytag 2011) and the
solar abundances of Asplund el al. (2009). Consider first the case of
an unspotted star; the appropriate curve of BC$_K$ versus colour is
calculated with temperatures and gravities corresponding to $t_{\rm
  95}$ for each mass track.  This is shown as a solid black line in Fig.~7a.

Also shown in Fig,~7a is a BC$_K$ versus colour curve at a fixed age of
1\,Gyr which approximates to the ZAMS. As expected this shows similar
values of BC$_K$ at bluer colours and higher mass stars but is
discrepant at lower masses. The reason is presumably that $t_{95}$ for
the higher mass stars is closer to their time of arrival on the ZAMS
than for lower mass stars, and hence their gravities are more similar
for a given $I-K$. The maximum discrepancy though is only 0.03\,mag at
0.07M/M$_{\odot}$ which would introduce systematic errors of only
$\sim$ 1.5 percent in the LDB age estimates for older ($>$100\,Myr)
clusters. This systematic error can be minimised by using a bolometric
correction curve at an age more appropriate to the expected LDB age of
the population. For example the curve for a fixed age of 150\,Myr (see
Fig.~7a) is virtually indistinguishable from one in which BC$_K$ is
calculated at $t_{\rm 95}$ at each mass\footnote{Or, if using an empirical
  bolometric correction relation derived from main sequence stars, a
  small theoretical correction, given by the gap between the 150\,Myr
  and 1\,Gyr curves in
  Fig.~7a, could be applied.} .

\subsection{The effect of starspots on colour magnitude curves}
Starspots affect the magnitude and colour of stars on the Hayashi
track in two ways; 
\begin{itemize}
	\item There is a small increase in the temperature {\it of the unspotted
          photosphere} of the spotted star, $T_s$ ($<4$ percent) 
          relative to that of an unspotted
          star, $T_u$ of the same age and luminosity. The magnitude of
          the change is found from the polytropic model as $(T_s/T_u)_{L,t} =
          (1-\beta)^{\tiny{-(1-B/4)/(A-B)}}$ (see equation 10 of Jackson
          \& Jeffries 2014). This makes the unspotted surface of a spotted 
          star slightly hotter (bluer)  than a totally unspotted star.
  \item There is also a contribution to the observed stellar flux from 
    the spotted area
    of the photosphere that always makes the star appear redder. The relative
    magnitude of this contribution depends on the spot temperature $T_{sp}$
    and the fraction of the surface covered by starspots,
    $\gamma=\beta/(1-T_{sp}^4/T_s^4)$
\end{itemize}

A simple two-temperature model is used to calculate the effect of
starspots on the bolometric correction at $t_{95}$, where
the $K$-band flux is the sum of the flux from the unspotted surface of
area
$(1-\gamma)4\pi R^2$ and temperature $T_s$, and the flux from the spotted
surface of area $\gamma 4\pi R^2$ and temperature $T_{sp}$. This gives a combined
$K$-band bolometric correction of
\begin{equation}
	{\rm BC}_{K,{\rm star}} = 2.5\log \left(\frac{1-\gamma}{1-\beta}10^{\tfrac{{\rm BC}_{K,T_s}}{2.5}}+	
	\frac{\gamma(\tfrac{T_{sp}}{T_s})^4}{1-\beta}10^{\tfrac{
            {\rm BC}_{K,T_{sp}}}{2.5}}\right)\, ,
\end{equation}
where BC$_{K,T}$ is the bolometric correction at temperature $T$ 
for an unspotted star at the Li depletion age.  A similar expression is used to evaluate
the $I$-band bolometric correction in order to determine BC$_K$ as a
function of colour.

The red dashed line in Fig.~7a shows BC$_K$ as a function of $I-K$
for a spotted star with $\beta$=0.2 and $T_{sp}/T_s$=0.8. At higher masses
the $K$-band flux from the spotted area reddens the star by $\sim
0.1$\,mag, but at
the same time BC$_K$ is increased by $\sim 0.05$\,mag such that the
spotted star still lies on the original BC$_K$ versus $I-K$ curve for
unspotted stars. At lower masses the $K$-band flux from the spotted areas is
much smaller relative to flux from the unspotted area, 
such that the presence of starspots induces a slight
blue shift due to the increase in temperature of the unspotted
photosphere, but again a compensatory change in BC$_K$ means the
relationship between BC$_K$ and $I-K$ is hardly changed.

Fig.~7b shows the difference between the bolometric correction
estimated from stellar colour (neglecting spot coverage) and the model
value from equation 6 for varying levels $\beta$ and $T_{sp}/T_s$. In
general the difference in values of BC$_K$ increases with $\beta$ but
peaks for values of $T_{sp}/T_s$ between 0.8 and 0.85 when the flux
from the spotted area is significantly reddened but still makes a
significant contribution to the stellar flux.
The results show that bolometric corrections
inferred from $I-K$ colour neglecting the effects of starspots
are systematically overestimated by just $\sim 0.01$\,mag for spot
coverage $\beta=0.3$. This corresponds to a systematic error of
only $\sim -0.7$ per cent in Li depletion age. A
similar analysis of BC$_K$ versus $V-K$ colour shows an expected
systematic error of $\sim -0.03$ in bolometric correction which
corresponds to just $\sim +1.4$ per cent error in LDB age. We regard these
systematic uncertainties as negligible given that $\beta = 0.3$ will
lead to changes in LDB age of $\sim +20$ per cent according to the
results of section 2.3. 

A more general treatment of this effect to include the many
ways that bolometric corrections could be estimated using different
photometric systems, or by using relationships between spectral type and
bolometric correction, is beyond the scope of this paper. The details
will depend exactly on which bands are used, how spectral types are
determined and how the weaker flux from spotted regions contributes
to the considered indicator.

\section{Discussion}

\subsection{The effects of starspots on LDB determinations and LDB ages}

The main result of this paper is that ages estimated by the LDB method
for spotted stars are systematically older by a factor of $(1-
\beta)^{-E}$ compared with estimates using theoretical models for
unspotted stars. The size of the effect depends on spot coverage
$\beta$ and the slope of the relationship between $\tau_{\rm LDB}$ and
$L_{\rm LDB}$, characterised by the index $E$ and illustrated in
Fig.~6. This {\em relative} age increase should apply whatever
theoretical models of unspotted stars 
are used to estimate the LDB age and the values of
$E$ will be similar too, because the relationship between $\tau_{\rm
  LDB}$ and $L_{\rm LDB}$ is almost model independent (e.g. Jeffries \&
Naylor 2001; Burke 2004). It should also apply for any choice of Li
depletion factor to define the LDB.  We have shown that the value of
$E$ is robust to the details of the polytropic models and that working
in terms of observed absolute magnitudes and bolometric corrections
does not change the basic result.  Hence the main uncertainty in
evaluating the significance of the effect is the value of $\beta$.

PMS stars at the LDB have a narrow range of spectral types from about
M4 at 20\,Myr (e.g. Binks \& Jeffries 2014) to M7 at 130\,Myr
(e.g. Cargile, James \& Jeffries 2010). Levels of magnetic activity in
such stars appear to be governed by a rotation-activity connection, and
as the spin-down timescales for stars at this mass are longer than
their LDB ages, it is likely that most stars at or around the LDB will
be rapidly rotating and highly magnetically active (e.g. Reiners \&
Basri 2008). This is confirmed from the short rotation periods inferred
from light curve modulation seen in a large fraction of M-type PMS
stars at ages of 10--150\,Myr (e.g. Irwin et al. 2007, 2008; Messina
et al. 2010) and from their ``saturated'' levels of chromospheric and
coronal activity (Messina et al. 2003; Jackson \& Jeffries 2010;
Jeffries et al. 2011).  The same rotational modulation provides
indirect evidence for spot coverage on PMS M-stars.  We are not aware
of Doppler imaging results for the mid-M or cooler objects
at the LDB in this age range, but results for young, early
M-dwarfs also provide ample evidence for extensive spot coverage
(e.g. Barnes \& Collier Cameron 2001). For older (150\,Myr) clusters
and the latest spectral types of stars at the LDB, it is possible that
levels of magnetic activity and consequent spot coverage begin to
decline (Reiners \& Basri 2008, 2010), although stars at the LDB in the
Pleiades (125\,Myr) and Blanco 1 (132\,Myr) are still
chromospherically active (Stauffer et al. 1998; Cargile et al. 2010).

\nocite{Messina2003a}
\nocite{Jeffries2011a}
\nocite{Jackson2010a}

The appropriate value of $\beta$ for active main sequence stars, let
alone PMS stars, is a matter of debate. The evidence is reviewed
extensively by Jackson \& Jeffries (2013, 2014) and Feiden \& Chaboyer
(2014). Rotational modulation of light curves gives only a lower limit
to $\beta$ -- any axisymmetric spot coverage will not contribute -- but
a $\beta$ of at least 0.1 is probable in the most active stars based on
the upper envelope of their photometric amplitudes (e.g. Messina,
Rodon\`o \& Guinan 2001; Messina et al. 2003).  Doppler imaging is also
likely to yield lower limits to spot coverage due to insensitivity to
axial symmetry and limited angular resolution (Solanki \& Unruh
2004). Spectroscopic constraints on spot coverage from modelling the
TiO bandheads and spectral energy distributions of active main sequence
G- and K-dwarfs suggest spot coverage in the range 20--50 per cent and
spotted to unspotted photospheric temperature ratios of 0.65-0.80,
leading to $0.1< \beta <0.4$ (O'Neal, Neff \& Saar 1998; Stauffer et
al. 2003; O'Neal et al. 2006). Whether such values can be extrapolated
to younger and lower mass PMS stars is uncertain.

\nocite{Messina2001b}

Indirect estimates of $\beta$ for low-mass PMS stars can
come from assuming starspots are responsible for their larger radii 
compared with models for inactive PMS stars of similar mass and age. 
This was the approach adopted by Jackson \& Jeffries (2014) who showed
that the radii of fully convective PMS stars
are increased by a factor $(1 - \beta)^{-0.45}$
at a given luminosity. From radius estimates for PMS M-dwarfs in NGC~2516, at an
age of 150\,Myr, they inferred $\beta \sim 0.5$. This value should
probably be regarded as an upper limit; Macdonald
\& Mullan (2013) have shown that the inclusion of magnetic inhibition of
convection in addition to dark starspots reduces the required values of
$\beta$.

In summary, the value of $\beta$ is quite uncertain (with an
uncertainty that grows with decreasing mass and hence older LDB ages)
but probably lies in the range 0.1--0.4 for the types of star
considered here. At the lower end of this range the presence of
starspots represents a small, but systematic increase of $\sim 5$ per
cent to LDB ages (slightly less for clusters older than 80\,Myr,
because $E$ is smaller -- see Fig.6). To put this in context, even this
correction is similar in size to making factors of two adjustment to
the mixing length adopted by evolutionary models or making quite
drastic changes to the radiative 
opacities or assumed boundary conditions of the outer
atmosphere (see figs. 3--5 in Burke et. al. 2004). Obviously if the
spot coverage were at the upper end of the range suggested above, then
LDB ages could be systematically increased by as much as 30 per cent;
a far larger correction than those due to any of the factors considered by
Burke et al. (2004). 

Any systematic increase in LDB ages would alter the conclusions of the
comparison with ages determined from fitting upper main sequence stars
and the nuclear turn-off in the HR diagram. The introduction of further
convective core overshooting or extra mixing due to rotation would be
required in order to those ages into concordance with the revised LDB
ages (see the discussions in Stauffer et al. 1998; Jeffries et
al. 2013; Soderblom et al. 2013).  A further effect of spots might be
to explain some of the ``blurring'' of the LDB that has been noted in
some clusters (e.g. Jeffries \& Oliveira 2005; Jeffries et
al. 2013). In principle, the LDB should be sharp, with little
difference in luminosity between those stars exhibiting no lithium and
those of slightly lower mass with their undepleted initial
abundance. In practice, there are stars with Li intermingled with
Li-depleted stars along the cluster sequence in a colour-magnitude
diagram. In section 3.2 we showed that spotted stars have virtually the
same relationship between bolometric correction and colour as unspotted
stars. Thus a coeval sequence of stars in the colour-magnitude diagram
is still a sequence of stellar luminosities even if some stars are more
spotted than others. However a star with larger $\beta$ will reach the
LDB (and hence be Li-depleted) at higher luminosity and hence brighter
absolute magnitude. Conversely, we expect Li-rich spotted stars to
coexist in the colour-magnitude plane with Li-poor stars with lower
spot coverage that have already reached the LDB. The size of any
intermingling region will be $\sim 2.5 \log (1 - \Delta \beta)$ mag,
where $\Delta \beta$ is the range of $\beta$ values.

\nocite{Jeffries2013a}
\nocite{Jackson2013a}
\nocite{Cargile2010a}
\nocite{Barnes2001a}
\nocite{Reiners2008a}
\nocite{Reiners2010a}
\nocite{Macdonald2013a}
\nocite{Solanki2004a} 
\nocite{Irwin2008a}
\nocite{Irwin2007a} 
\nocite{Messina2010a}

\subsection{LDB ages and other magnetically caused structural changes} 

The analysis in Sections 2 and 3 considers only the effect of starspots
on the LDB age. However, other authors have considered a
number of ways that the presence of dynamo-generated magnetic fields
might alter the structure of low-mass stars. The degree to which
magnetic fields affect LDB age estimates depends primarily on
how they effect the rate of heat loss from the star and hence the rate
of change of radius with time. Broadly speaking there are two ways in
which this can happen (e.g. Chabrier, Gallardo \& Baraffe 2007). The first
is a reduction in the efficiency of radiative heat transfer from the
surface of the star due to the presence of starspots, which we have
discussed here and which can be characterised with $\beta$, the
fraction of flux blocked from leaving the star (Spruit
1982). The second is a reduction in the efficiency of convective heat
transfer in the interior of the star, which increases the radius of the
star at a uniform surface temperature. This may be characterised as a
reduction in mixing length (Somers and Pinsonneault \& 2014), or by
changes to the Schwarzschild stability criterion (Mullan \& Macdonald
2001, Feiden \& Chaboyer 2014).

Somers \& Pinsonneault (2014) claimed that a reduction in mixing length
would increase LDB ages and this was numerically established by Burke
et al. (2004). Here we can briefly see how this result fits into our
polytropic framework and show that the increase in LDB age can be more generally
expressed in terms of an increase in radius at a given age rather than
in terms of $\beta$. By equating the rate of change of gravitational
potential energy to its luminosity, for a polytropic star (of index $n$) 
contracting along a vertical Hayashi track, 
we can write down the time taken to reach a particular
radius (in this case the radius at which Li is depleted)
\begin{equation}
t_{\rm Li}= \frac{1}{10-2n}\frac{GM}{4\pi \sigma T_{\rm
    eff}^4}\frac{1}{R_{\rm Li}^3}
\end{equation} 
We now say that the radius of a star at a given age is magnetically inflated 
by a factor $(R_i/R_u)_t$ where suffix $i$ refers to
the inflated star and suffix $u$ the reference star. If we further 
assume that the surface temperature is reduced by a factor
$(T_i/T_u)_t = (R_i/R_u)_{t}^{-1/2}$, such that $M/L$ is unchanged, 
then the Li depletion age is increased by a factor $t_{i,{\rm Li}}/
t_{u,{\rm Li}}
= (R_i/R_u)_{t}^{2}$ and the luminosity decreased by a factor
$L_{i,{\rm Li}}/
L_{u,{\rm Li}} = (R_i/R_u)_{t}^{-2}$. These two factors are equivalent to the
shifts of $(1- \beta)^{-1}$ and $(1- \beta)$ established for spotted
stars in section 2.3 provided that  $(R_i/R_u)_t \simeq (1
-\beta)^{-0.5}$. But that is precisely what was found in Jackson \& Jeffries
(2014), where we showed that $(R_i/R_u)_{L,t} = (1 -\beta)^{-0.5}$ 
if the Hayashi tracks are vertical. Hence, although this
is a very simple model (the Hayashi tracks are not vertical and the
polytropic index might be changed depending on the radial profile of
magnetic fields -- e.g. Feiden \& Chaboyer 2014), it shows that if
radii are inflated by a given factor, the onset of
Li-burning will be delayed and LDB ages increased by a similar amount
irrespective of the exact mechanism(s) causing the radii to be larger.

Whilst there might be little difference in $L_{\rm LDB}$ for stars of a
given age that are made larger by either starspots or a reduction in
convective efficiency, there {\it is} a key observational distinction.
Starspots change the bolometric correction and colours of stars in
compensatory ways so that the colour and magnitude of a star at the LDB
closely mimics those of an unspotted star of slightly different mass. On
the contrary, stars that are made larger but have a uniform surface
temperature will be significantly redder than unspotted stars of
similar luminosity or magnitude. The reddening with respect to
non-magnetic stars grows (e.g. for radii
increased by 11 per cent) from 0.12
mag in $I-K$ at $M=0.5\,M_{\odot}$ ($\tau_{\rm LDB} \simeq 20$\,Myr) to
0.35 mag at $M=0.075\,M_{\odot}$ ($\tau_{\rm LDB} \simeq
150$\,Myr). Thus the {\it colour} of the LDB for inflated stars with a
uniform photospheric temperature would be redder than if the inflation
were caused by dark photospheric inhomogeneities. The difference
should be quite marked at the older end of the age range considered if
radii were increased to this extent.

\nocite{Chabrier2007a}
\nocite{Oneal1998a}
\nocite{Oneal2006a}
\nocite{Stauffer2003a}

\section{Summary}

Polytropic models of fully
convective PMS stars are used to investigate the effects of
cool starspots on ages determined from the Lithium
Depletion Boundary (LDB). The effect of spots is to slow the rate of
descent along Hayashi tracks leading to lower core temperatures at a
given age or luminosity. This delays the onset of Li destruction and
means that if the luminosity at which Li is depleted can be measured --
the LDB -- then the age inferred from this should be increased. 
Our specific findings are:
\begin{itemize}
\item The age, $\tau_{\rm LDB}$, inferred from the luminosity of the
  LDB, $L_{\rm LDB}$,  is {\it systematically} increased by a
  factor $(1- \beta)^{-E}$ compared with an age estimated from
  unspotted evolutionary models. Here, $\beta$ is the equivalent
  coverage of dark starspots and the exponent $E$  is related to the 
  rate of change of  $\tau_{\rm LDB}$ with $  L_{\rm LDB}$; $E \simeq 1 + d\log \tau_{\rm LDB}/d\log L_{\rm LDB}$. For ages $<80$\,Myr, $E \sim 0.5$ and decreases  towards $\sim 0.3$ at older ages as shown in Fig.~6.
\item We also find that spotted stars show virtually the same
  relationship between $K$-band bolometric correction and colour as unspotted
  stars, so that estimates of age based on the absolute $K$ magnitude
  will be affected in exactly the same way.
\item The appropriate value(s) of $\beta$ to adopt are highly
  uncertain, but may lie in the range 0.1--0.4. Even at the lower end of
  this range, the systematic shift in LDB ages is comparable with the
  largest theoretical uncertainties previously claimed for the
  technique (e.g. Burke et al. 2004). For the largest values of $\beta$
  considered, LDB ages would be increased by 30 per cent (16 per
  cent at older ages), a significant change that would require
  re-evaluation of the physics in higher mass stellar
  models if ages inferred from these were to agree with LDB ages.
\item We note that any magnetic
  effect that causes PMS stellar radii to be larger at a given age, 
  whether that be starspots or a reduced convective efficiency,
   will produce similar increases in inferred LDB ages. However,
   increased radii with uniform photospheric temperatures due to
   suppressed convection in the interior will cause
   the LDB to be observed at significantly redder colours than for
   spotted stars of equivalent radii, or for non-magnetic stars. 
\end{itemize}

\section*{Acknowledgments}

RJJ and RDJ wish to thank the UK Science and Technology Facilities Council for
financial support.

\bibliographystyle{mn2e} 
\bibliography{references}


\bsp 

\label{lastpage}

\end{document}